\documentstyle[aps,epsfig]{revtex}

\begin{document}

\title{Luttinger model approach to interacting one--dimensional fermions
in a harmonic trap}

\author{W. Wonneberger} 

\address{Abteilung f\"ur Mathematische Physik, Universit\"at Ulm, D89069 Ulm, Germany}

\date{\today}

\maketitle

\begin{abstract}
A model of interacting one--dimensional fermions confined to a 
harmonic trap is proposed. The model is treated analytically
to all orders of the coupling constant by a method analogous to that
used for the Luttinger model. As a first application, the particle
density is evaluated and the behavior of Friedel oscillations under 
the influence of interactions is studied. It is found that 
attractive interactions tend to suppress the Friedel oscillations
while strong repulsive interactions enhance the Friedel oscillations
significantly. The momentum distribution function and the relation
of the model interaction to realistic pair interactions are also 
discussed.
\end{abstract}


\section{Introduction}

The rapid progress in cooling and magnetic trapping technology 
\cite{TEC95,LRW96,VFP98,FGZ98,DCS99,TOZ99,RHH99} 
makes it conceivable to produce a neutral ultra--cold  
quantum gas of quasi one--dimensional degenerate fermions following the
recent achievement in three spatial dimensions \cite{DMJ99}. Some
experimental aspects of such a quasi one--dimensional fermion 
gas are discussed in \cite{GWSZ00}.

In three spatial dimensions, identical spin polarized fermions experience only 
a weak interaction because the s--wave scattering
contribution from short ranged pair potentials is forbidden. Scattering 
processes involving higher angular momenta have energy thresholds and freeze
out at temperatures below the Fermi temperature $T_F$ \cite{DM99}.
At zero temperature only a weak but long ranged dipole--dipole 
interaction \cite{GRP00,GER01} remains.
 
This classification scheme does not apply to fermions in one spatial dimension. 
One can only state that a contact interaction has no effect. It is tempting to
completely neglect all interactions and treat the one--dimensional fermion system as 
an ideal gas in a harmonic trap. Some exact single particle properties of this
system were given in \cite{VMT00,GWSZ00}.
The theory of Luttinger liquids \cite{T50,L63,ML65,LP74,H81} 
(for reviews cf. \cite {E79,V95,Sch95}) suggests,
however, that even weak interactions destroy the Fermi liquid picture in one
spatial dimension. This holds also true for bounded Luttinger liquids 
\cite{C84,EA92,FG95,EG95,WVF96,MEJ97,VYG99} which display algebraically decaying 
correlations. The confining potential for the cases studied so far is provided 
by hard walls. In ultra--cold quantum gases a harmonic confining potential is 
more appropriate. 

Besides the Luttinger approach to confined one--dimensional 
fermions, a wealth of exact results exist for the Calogero--Sutherland model 
\cite{C69,S71} with harmonic confinement and its extensions (cf., e.g., \cite{KK94}). 
These results hold
for the $1/r^2$ pair interaction. A recent path integral approach to the thermodynamics 
of harmonically confined fermions \cite{BDL98} assumes a harmonic pair interaction.
A different treatment of that model is given in \cite{ZGOM00} and includes 
results for Green's functions of one-dimensional fermions. The present approach, 
which is expected to be asymptotically correct for large fermion numbers, is more 
flexible with respect to the form of the interaction.

The method proposed is a transcription of the
elementary solution of the Luttinger model (cf., e.g., \cite{Sch95}) to the present problem.
It requires, however, a different bosonization scheme which we adopt from \cite{SchM96}.

As a first application, we calculate the particle density. The particle density depends
perturbationally on the interaction strength. But we are still able to present some
interesting interaction effects related to the Friedel \cite{F58} oscillations. 

The paper is organized as follows: In Sec. 2, we present the basic arguments of our
approach. Sec. 3 discusses the occupation probabilities of single particle states
and the corresponding particle density for two interaction models. Sec. 4 shortly 
discusses the momentum distribution function and its relation to the particle density. 
In an Appendix, the pertinent question of the relation between the model interaction 
and realistic interactions is studied and estimates of their respective strengths 
are made.

\section{Luttinger Approach}

We consider a gas of spinless identical fermions in one 
spatial dimension and trapped in a harmonic potential

\begin{equation}\label{1.1}
V(z) = \frac{1}{2} m \omega ^2 _{\ell} z^2.
\end{equation}

The Hamiltonian in second quantization but without interactions is

\begin{equation}\label{1.2}
\hat{H}_0 = \sum^\infty _{n=0} \hbar \omega _n\,\hat{c}^+_n \hat{c}_n,
\end{equation}

with one--particle energies 
 
 \begin{equation}\label{1.2a}
 \hbar \omega _n = \hbar \omega _{\ell} (n+1/2),\,\,\, n=0,1,... .
 \end{equation}

 The fermion creation operators 
 $\hat{c}^+$ and destruction operators $\hat{c} $ obey the fermionic algebra
 $ \hat{c}_m \hat{c}_n^+ + \hat{c}_n^+ \hat{c}_m = \delta _{m,n}$.   
 This ensures that each (non-degenerate) energy level $\epsilon _n = \hbar \omega _n $ with 
 (real) single particle wave function
 
 \begin{equation}\label{1.3}
 \psi _n(z) = \sqrt{\frac{\alpha}{2^nn!\pi^{1/2}}} 
 \,e^{-\alpha^2 z^2/2 }\,H_n (\alpha z)
 \end{equation} 

 is at most singly occupied. The
 intrinsic length scale of the system is the oscillator length $l = \alpha ^{-1}$ 
 where $\alpha$ is defined by $\alpha^2=m \omega _\ell/\hbar$. 
 $ H_n$ denotes a Hermite polynomial.
 
 The spatial extension of the bounded system of $N$ fermions is measured
 by the Fermi width $L_F$, i.e., half the extension of the classically allowed
 region at the Fermi energy:

 \begin{equation}\label{1.4}
 L_F = \frac{1}{\alpha} \sqrt{2N-1} \equiv L_{n=N-1}.
 \end{equation}

 The Fermi energy itself is given by:

\begin{eqnarray}\label{1.5}
\epsilon _F = \hbar \omega _\ell (N-1) + \frac{1}{2} \hbar
\omega _\ell =  \hbar \omega _\ell (N-\frac{1}{2}),
\end{eqnarray}

with an associated Fermi wave number \cite{BR97,GWSZ00}

\begin{eqnarray}\label{1.6}
 k_F \equiv \alpha \sqrt{2 N - 1} \equiv \sqrt{2m\epsilon _F/\hbar^2}.
\end{eqnarray}

The density oscillations well inside the trap can be identified \cite{GWSZ00} as
Friedel oscillations \cite{F58} with wave number $2 k_F$.
 
As is seen from eq. (\ref{1.2a}), the single particle energies depend linearly
on the quantum number $n$. A linear dispersion is one of the 
requirements which allow the treatment of the Luttinger model. Another 
requirement is the presence of an
anomalous vacuum with linear dispersion extending to arbitrary negative energies. 
In our case, these fictitious free states belong to $n<0$. Formally, they have wave 
functions $\psi _p (z) \propto {\cal{D}} _p \left( \sqrt{2}\alpha z \right)$
with $p=-n<0$, ($n$: positive integer, ${\cal{D}}$: parabolic cylinder function)
which do not belong to the Hilbert space. This deficiency is likely to have
no effect because these functions are still normalizable in any finite
spatial region $[-L,L]$ with $L \gg L_F$ to which all finite range interaction 
effects are confined. Furthermore, one expects that for sufficiently large
fermion number $N$ the presence of the anomalous vacuum is harmless for 
processes near the Fermi energy $\epsilon _F \propto N$ \cite{H81}.
 
Exploiting the presence of the anomalous vacuum, it is easy to show
that the density fluctuation operators

\begin{eqnarray}\label{1.8}
 \hat{\rho} (m) \equiv \sum _p \hat{c}^+_{p+m} \hat{c}_p 
\end{eqnarray}

obey bosonic commutation relations

\begin{eqnarray}\label{1.9}
 [\hat{\rho} (-p), \hat{\rho} (q)] = p\, \delta _{p,q}
\end{eqnarray}

for all integers $p$ and $q$. The bosonic operators defined according to

\begin{eqnarray}\label{1.10}
\hat{\rho} (p) = \left \{
\begin{array}{lll}
\sqrt{|p|} & \hat{d}_{|p|}, & p< 0,\\
\sqrt{p}   & \hat{d}^+_p,   & p> 0,
\end{array} \right. 
\end{eqnarray}

thus satisfy canonical commutation relations:
 
\begin{eqnarray}\label{1.11}
[\hat{d}_m, \hat{d}^+_n ] = \delta _{m,n}
\end{eqnarray}

for positive integers $m$ and $n$.
The following argument is taken over from the Luttinger model: Since the 
free Hamiltonian $\hat{H}_0 $ has the same commutators with $\hat{\rho}(p)$:

\begin{eqnarray}\label{1.13}
  [ \hat{H}_0, \hat{\rho}(p) ] = \hbar \omega _\ell \,p\,
\hat{\rho}(p), 
\end{eqnarray}

as

\begin{eqnarray}\label{1.14}
 \tilde{H}_0 = \hbar \omega _\ell \sum _{m>0}  m \,\hat{d}^+_m \hat{d}_m
\end{eqnarray}

has with the corresponding $\hat{d}_m$ and $\hat{d}^+_n$, the low energy (density
wave) excitations of the fermionic Hamiltonian can as well be described 
in the bosonic Hilbert space.

The form of the admissible four--fermion interaction in the full
Hamiltonian is dictated by the requirement that it can be expressed in terms of
density fluctuation operators. This restricts the form of the matrix elements of
the interaction operator

\begin{eqnarray}\label{1.15}
 \hat{V} =\frac{1}{2} \sum _{mnpq}
V(m,p;q,n)\,(\hat{c}^+_m \hat{c}_q ) ( \hat{c}^+_p \hat{c}_n ).
\end{eqnarray}

Two solvable forms were found: The matrix elements

\begin{eqnarray}\label{1.16}
V (m,p;q,n) = \int _{-\infty}^\infty d1d2d3d4\,\psi _m (1) \psi _p (2) V(1,2;3,4)
\psi _q (3) \psi _n (4)
\end{eqnarray}

must simplify according to

a)
\begin{eqnarray}\label{1.17}
V (m,p;q,n) = V _a (|q-m|) \,\delta _{ m-q, n-p}\,, 
\end{eqnarray}

which leads to

\begin{eqnarray}\label{1.18}
\hat{V}_4 = \frac{1}{2} \sum _{m>0} V _a (m) \,m \left\{ \hat{d}^+_m
\hat{d}_m +\hat{d}_m  \hat{d}^+_m \right\}, 
\end{eqnarray}

and
 
b)
\begin{eqnarray}\label{1.19}
  V (m,p;q,n) = V _b (|q-m|)\,  \delta _{q-m, n-p}\,, 
\end{eqnarray}

which gives

\begin{eqnarray}\label{1.20}
\hat{V}_2 =  \frac{1}{2} \sum _{m>0} V _b (m) \,m \left\{ \hat{d}^2_m
+\hat{d}^{+2}_m  \right\}. 
\end{eqnarray}

$V_a(m)$ and $V_b(m)$ correspond to the $g_4$ and $g_2$ coupling functions of 
the Luttinger model.
The Appendix discusses the relation of the present interactions to realistic 
four--fermion interactions.

The total bosonic interaction operator, which is added to the free Hamiltonian
 (\ref{1.14}), thus is

\begin{eqnarray}\label{1.21}
 \hat{V} =  \frac{1}{2} \sum _{m>0} m V _b (m)\left\{ \hat{d}^{+2}_m
+ \hat{d}^2_m  \right\}
+ \frac{1}{2} \sum _{m>0} m V _a (m)\left\{ \hat{d}^+_m	\hat{d}_m +
\hat{d}_m \hat{d}^+_m  \right\}. 
\end{eqnarray}

It is diagonalized in the bosonic Hilbert space by the Bogoliubov transformation

\begin{eqnarray}\label{1.22}
\tilde{H} = \tilde{H} _0 + \hat{V} =
\sum _{m>0} m\,\epsilon _m \hat{f}^+_m \hat{f}_m +\mbox{const.}, 
\end{eqnarray}

with the unitary operator

\begin{eqnarray}\label{1.23}
 \hat{S} = \exp \left\{ \frac{1}{2} \sum _{m>0} \zeta _m
\left( \hat{f}^2_m - \hat{f}^{+2}_m \right) \right\}. 
\end{eqnarray}

This transformation relates the $\hat{d}$--operators to new canonical 
$\hat{f}$--operators according to

\begin{eqnarray}\label{1.24}
\hat{d} _m = \hat{S}^+ \hat{f} _m \hat{S} = \hat{f} _m \cosh \zeta _m
-\hat{f}^+ _m\sinh \zeta _m. 
\end{eqnarray}

The transformation parameters $\zeta _m$ are given by

\begin{eqnarray}\label{1.25}
\tanh 2 \zeta _m = \frac{V_b (m)}{\hbar \omega _\ell + V_a (m) }. 
\end{eqnarray}

They depend also on the sign of the interactions.
The spectrum of the density wave excitations is found to be

\begin{eqnarray}\label{1.26}
 \epsilon _m = \sqrt{ \left( \hbar \omega _\ell + V_a (m) \right)^2
- V^2_b (m) } 
\equiv  \left( \hbar \omega _\ell + V_a (m) \right)
\cosh 2 \zeta _m - V_b (m) \sinh 2 \zeta _m. 
\end{eqnarray}

It is also seen that $V_a(m) \neq 0$ only renormalizes the single particle energy
$\hbar \omega _\ell$. For large $N$ and at the Fermi edge, a local pair potential 
contributes equally to $V_a(m)$ and $V_b(m)$ because of its symmetry $m 
\leftrightarrow q$ or $n \leftrightarrow p$ (see Appendix). 
 
The matrix elements $V_a$ and $V_b$ determine a set of coupling constants 
 
\begin{eqnarray}\label{1.27}
\gamma _m \equiv\sinh^2 \zeta _m = \frac{1}{4} \left( K_m + \frac{1}{K_m} - 2 \right)\ge 0,
\end{eqnarray}

where $K_m$ is given by

\begin{eqnarray}\label{1.28}
 K_m = \sqrt {\frac{\hbar \omega _\ell + V_a (m) - V_b (m)}
{\hbar \omega _\ell + V_a (m) + V_b (m)} } = \exp(-2 \zeta _m).
\end{eqnarray}

This is in complete analogy to the Luttinger model. There is an obvious consistency 
condition

\begin{eqnarray}\label{1.29}
 |V_b (m) | \le | \hbar \omega _\ell + V_a (m) |, 
\end{eqnarray}

and a more subtle stability condition \cite{H81}:

\begin{eqnarray}\label{1.30}
\sqrt{m}\, \frac{V_b (m)}{\hbar \omega _\ell + V_a (m)}
\stackrel{m \rightarrow \infty}{\longrightarrow} 0. 
\end{eqnarray}

We will mostly assume $V_a(m)=V_b(m)\equiv V(m)$. Then the simple relations 

\begin{eqnarray}\label{1.31}
\epsilon _m = \frac{\hbar \omega _\ell}{K_m},\quad V(m)=\frac{1}{2}\,\left(
\frac{1}{K_m^2}-1\right)\,\hbar \omega _\ell
\end{eqnarray}

hold. We will also consider two specific interaction models:

\begin{itemize}
\item[1.] A toy model, called IM1, which is defined by $V(m)=V(1)(\delta _
{m,1}+\delta _{m,-1})$ with a real amplitude $V(1)$.
\item[2.] The analogue of the usual Luttinger liquid interaction with a small 
exponential decay
in the coupling constants $\gamma _m$ according to $\gamma _m = \exp(-r_\gamma m)
\,\gamma _0$ in order to satisfy (\ref{1.30}).
In the particle density considered below, further coupling constants $\alpha _m$
(cf. (\ref{2.5a})) appear. For these coupling constants 
we assume a decay according to $\alpha _m=\exp(-r_\alpha m/2)\,\alpha _0$
with $\alpha _0 =sgn(V(1))\sqrt{\gamma _0(1+\gamma _0)}$. We call this model 
IM2. This model is not fully consistent because the exponential decay of 
$\gamma _m$ induces a non--exponential decay of $\alpha _m$ unless $\gamma _0
\ll 1$ or $\gamma _0 \gg 1$ when $r_\alpha \rightarrow r_\gamma$ or $r_\alpha \rightarrow 
2 r_\gamma$, respectively. By choosing $r _\alpha$ properly, it is possible to restrict 
the error. The choice $r_\gamma = r_\alpha \propto \sqrt{N^{-1}}$, which corresponds
to the length scale $L_F$, also poses no problem when $N$ is large. 
\end{itemize}

An essential step in the applicability of the Luttinger model is the correspondence
between fermionic creation and destruction operators and exponents of 
bosonic fields involving the $\hat{d}$ (or $\hat{f}$)--operators 
\cite{LP74,H81,E79,V95,Sch95}. The standard form
of this bosonization is apparently not applicable to the present case.
Instead we will use a method initially introduced in \cite{SchM96}.
The authors define an auxiliary field according to

\begin{eqnarray}\label{1.32}
 \hat{\psi} _a (v) \equiv \sum^\infty _{l=-\infty} e^{ilv} \hat{c}_l
= \hat{\psi}_a (v + 2 \pi), 
\end{eqnarray}

and prove the bosonization formula for the bilinear combination

\begin{eqnarray}\label{1.33}
 \hat{\psi}^+_a (u) \hat{\psi}_a (v) = G_N ( u - v )
\exp \left\{ - i \left( \hat{\phi}^+(u) - \hat{\phi}^+(v) \right) \right\}
\exp \left\{ - i \left( \hat{\phi} (u) -\hat{\phi} (v) \right) \right\},     
\end{eqnarray}

involving the non--Hermitian bosonic field

\begin{eqnarray}\label{1.34}
 \hat{\phi}^+ (v) = i \sum^\infty _{n = 1} \frac{1}{ \sqrt {n} } e^{-inv}
\hat{d}^+ _n \neq \hat{\phi}(v),
\end{eqnarray} 

and the Fermi sum
 
\begin{eqnarray}\label{1.35}
\quad G_N (u) = \sum^{N-1} _{l = - \infty} e^{-i l (u+i \epsilon)}.
\end{eqnarray}
 
Equations (\ref{1.22}), (\ref{1.24}), (\ref{1.26}), and (\ref{1.32}) to (\ref{1.35}) 
provide a framework for the calculation of the fermion density and also of
correlation functions of the density for one--dimensional fermions in a harmonic trap 
interacting via (\ref{1.21}). They do not, however, allow to calculate the time
dependence in the single particle Green's function.

In the remaining part we apply the concept to the fermion density in the trap.

\section{Perturbed Particle Density}

The fermion density is given by the expectation value

\begin{eqnarray}\label{2.1}
n(z) = \langle \hat{\psi}^+ (z) \hat{\psi} (z) \rangle. 
\end{eqnarray}
 
Decomposing the fermionic destruction operator $\hat{\psi} (z)$ according to

\begin{eqnarray}\label{2.2} 
 \hat{\psi} (z) \equiv  \sum^\infty _{m=0}  \psi _m (z) \, \hat{c} _m, 
\end{eqnarray}
 
gives

\begin{eqnarray}\label{2.3}
 n (z) = \sum^\infty _{m,n =0} \psi _m (z) \psi _n (z) \,
\langle \hat{c}^+ _m \hat{c} _n \rangle. 
\end{eqnarray}

Thus, the computation of the expectation values $\langle \hat{c}^+ _m \hat{c} _n \rangle $
is required. They become non--diagonal in the presence of interactions. Using the 
prescription of the last section, we find

\begin{eqnarray}\label{2.4}
 \langle \hat{c}^+ _m  \hat{c}_n \rangle = \sum _{l=-\infty}^{N-1}
\int^{2 \pi}_0 \int^{2 \pi}_0 \,\frac{du dv}{4 \pi^2} \,e^{i(m-l)(u+i\epsilon) 
- i(n-l)(v-i\epsilon)}
\,\, \langle e^{-i \hat{\phi}^+ (u) + i	\hat{\phi}^+ (v)}
 e^{-i	\hat{\phi} (u) + i  \hat{\phi} (v)} \rangle. 
\end{eqnarray}
 
By standard methods (bosonic Wick theorem $\langle \exp[\hat{A}]\rangle = 
\exp[ \langle \hat{A}^2 \rangle/2]$ for any linear combination $\hat{A}$ of
$\hat{f}_m$ and  $\hat{f}_n^+$) and at zero temperature (when $\langle \hat{f}_m 
\hat{f}_n^+ \rangle = \delta _{m,n}$), the expectation value $\langle \quad
\rangle \equiv \exp[-W]$ is calculated from

\begin{eqnarray}\label{2.5}
 W= W(u,v) = 2 \sum^\infty _{m=1} \frac{1}{m} \,
\left[ \gamma _m - \alpha _m \,\cos m(u + v) \right]
 \left\{ 1 - \cos m (u - v) \right\}, 
\end{eqnarray} 

with

\begin{eqnarray}\label{2.5a} 
 \alpha _m  \equiv sgn(V(m)) \sqrt{\gamma _m(1+\gamma _m)}. 
\end{eqnarray}

In this way, a reduction to ''quadratures'' is achieved but nontrivial summations 
and integrations remain.

$W$ is a real and an even function of its arguments. This gives the symmetries

\begin{eqnarray}\label{2.6}
\langle \hat{c}^+ _m \hat{c} _n \rangle =\langle \hat{c}^+ _n \hat{c} _m \rangle
=\langle \hat{c}^+ _m \hat{c} _n \rangle^*.
\end{eqnarray}

It will also turn out that $\langle \hat{c}^+ _m \hat{c} _n \rangle$ vanishes unless
$|m-n|$ is even. The particle density thus admits the representation:

\begin{eqnarray}\label{2.7}
n (z,N,T=0) = \sum^\infty _{M = 0} \psi _{M}(z)^2\,\langle \hat{c}^+ _M \hat{c} _M \rangle
+2\sum^\infty _{M = 1} \sum _{p=1}^{M}\psi _{M-p}(z)\,\psi _{M+p}(z)\,
\langle \hat{c}^+ _{M-p}\hat{c} _{M+p} \rangle.
\end{eqnarray}

The $l$--summation in (\ref{2.4}) can be performed and the two integrations appropriately
transformed using the $2\pi$--periodicity of the integrand in (\ref{2.4}). 

Turning to the interaction model IM1, one of the integrations can be performed, giving

\begin{eqnarray}\label{2.8}
\langle \hat{c}^+ _{M-p}\hat{c} _{M+p} \rangle=\frac{1}{2}\,\delta _{p,0}
-\frac{1}{2\pi}\int _{-\pi}^\pi ds\,&&\left\{\frac{\sin((M+1/2-N)s)}{2\sin(s/2)}
\right\}\\[4mm]\nonumber
&&\exp[-2\gamma _1(1-\cos(s))]\,I_p(2\alpha _1(1-\cos(s))).
\end{eqnarray} 

Evidently, the factor in curly brackets, which stems from the $l$--summation, provides
particle--hole symmetry: For $p \neq 0$ it has the form

\begin{eqnarray}\label{2.9} 
\langle \hat{c}^+ _{2N-1-M-p}\hat{c} _{2N-1-M+p} \rangle
= -\langle \hat{c}^+ _{M-p}\hat{c} _{M+p} \rangle,
\end{eqnarray}

i.e., in a plot of $\langle \hat{c}^+ _{M-p}\hat{c} _{M+p} \rangle$ versus $M$ the point
$(N-1/2,0)$ is the point of inversion symmetry.
For $p=0$ one obtains the occupation numbers $\langle \hat{c}^+ _{M}\hat{c}_{M} \rangle$,
which satisfy 

\begin{eqnarray}\label{2.10}
\langle \hat{c}^+ _{2N-1-M}\hat{c} _{2N-1M} \rangle
=1-\langle \hat{c}^+ _{M}\hat{c}_{M}\rangle.
\end{eqnarray}

In the present fermionic case, they can be interpreted as occupation probabilities
$P(M)$ of the single particle state $\psi _m$. The occupation probabilities show 
the expected smearing of the Fermi edge around $M=N-1$ due to interactions.
The inversion symmetry point for the probabilities is $(N-1/2,1/2)$.

Several statements concerning (\ref{2.8}) can be made:

\begin{itemize}
\item[1.] The result of first order many body perturbation theory in fermionic Hilbert
space are exactly reproduced for $V_a(1)=0$ by expanding (\ref{2.8}) to first order 
in the coupling constant $V_b(1)$ (i.e., $\gamma _1 \rightarrow 0$, $\alpha _1 
\rightarrow \zeta _1 \rightarrow V_b(1)/(2 \hbar \omega _\ell)$) \cite{FG}.

\item[2.] The results of direct numerical diagonalization of the perturbed
fermionic Hamiltonian agree with the numerical evaluation of (\ref{2.8}) 
up to $|\alpha _0| =1$ and for particle numbers $N$ of the order of $10$, the 
range investigated so far \cite{FL00}.

\item[3.] The sum rule

\begin{eqnarray}\label{2.10a} 
\int _\infty^\infty\, dz\, n(z,N,T=0)=\sum _{M=0}^\infty \,\langle \hat{c}^+ _{M}
\hat{c} _{M} \rangle= N
\end{eqnarray} 
  
is exactly obeyed by (\ref{2.8}): Summing the expression in curly brackets in 
(\ref{2.8}) from $M=0$ to $Q \gg N$ leads to the result $2 \pi \delta(s)
((Q+1)/2 -N)$. 
    
\end{itemize}

Returning to the particle density, the modification of $P(M)$ and the appearance of
non--diagonal contributions has pronounced effects: For increasingly negative values
of $\alpha _1$ the Friedel oscillations are diminished while positive $\alpha _1$ 
increase their amplitude. This will be even more pronounced in the interaction model IM2
in which several basic interactions are superimposed.

The corresponding formula for $\langle \hat{c}^+ _{M-p}\hat{c} _{M+p} \rangle$ is more 
complicated and can be given in the form:

\begin{eqnarray}\label{2.11}
\langle \hat{c}^+ _{M-p}\hat{c} _{M+p} \rangle= \frac{1}{2}\, \delta _{p,0}
-&&\int^{\pi}_{-\pi}\frac{dt}{2\pi}\, \frac{\cos(p\,t)}{[1+Z_\alpha-\cos(t)]^{\alpha _0}}
\\[4mm]\nonumber
&&\int^{\pi}_{-\pi} \frac{ds}{2\pi}\,\left\{\frac{\sin((M+1/2-N)s)}{2\sin(s/2))}\right\}
\left(\frac{Z_\gamma}{[1+Z_\gamma-\cos(s)]}\right)^{\gamma _0}
\\[4mm]\nonumber
&&\quad\quad\quad\,\,[(1+Z_\alpha-\cos(t-s))(1+Z_\alpha-\cos(t+s))]^{\alpha _0/2}. 
\end{eqnarray}

The quantities $Z_\gamma$ and $Z_\alpha$ result from the decay functions
of the coupling constants $\gamma _m$ and $\alpha _m$. They are given by

\begin{eqnarray}\label{2.14}
Z_\gamma=\cosh(r_\gamma)-1,\quad Z_\alpha=\cosh(r_\alpha/2)-1.
\end{eqnarray} 

Again, the sum rule (\ref{2.10a}) is strictly fulfilled.

Figure 1 shows the corresponding particle density for $\alpha _0= \pm 1$, 
$r_\gamma =0.3$, and $r_\alpha =0.4$ obtained by numerical integration of 
(\ref{2.11}) and by summing the series (\ref{2.7}) up
to $M=2 N=20$. The occupation probabilities are smoothed out and more spectral 
weight is found above the Fermi energy. Correspondingly, the particle density 
outside the
classical region is enhanced. Most startling, however, is the strong enhancement 
of the Friedel oscillations in case of strong repulsive interaction and their nearly 
complete suppression for strong attractive interactions.

\begin{figure}[ht]
 \begin{center}
 \epsfig{figure=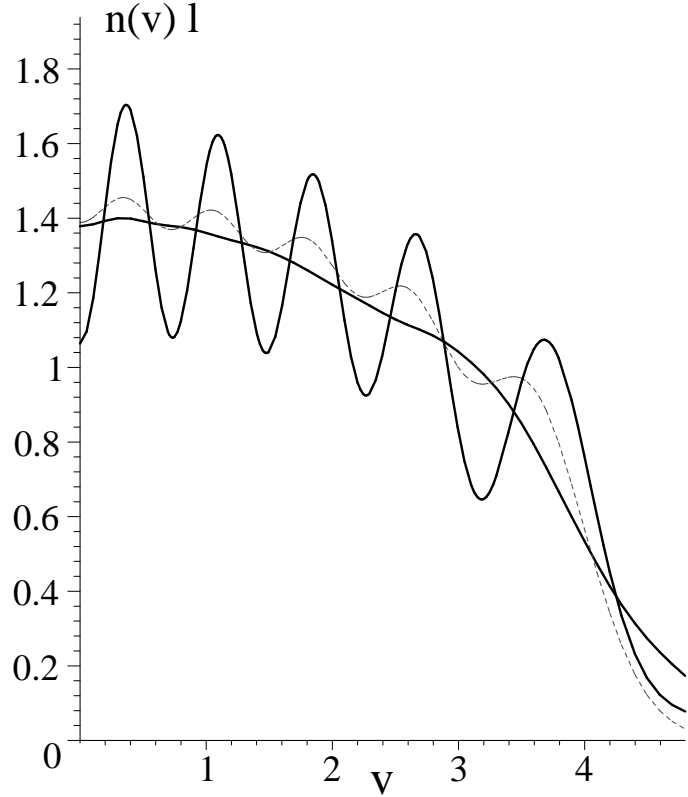,width=0.5\columnwidth}
 \end{center} 
 \caption{\small Dimensionless particle density $n$ in units of the inverse of 
 the oscillator length $l$
 versus dimensionless distance $v=z/l$ from the center of the one--dimensional harmonic 
 trap for $N=10$ interacting spinless fermions at zero 
 temperature. Broken curve shows unperturbed Friedel oscillations. Strongly oscillating 
 curve refers to a repulsive interaction with $\alpha _0=1$ while the smooth curve
 is for an attractive interaction with $\alpha _0=-1$. Interaction model 2 has been used. 
 }
\end{figure}
 
 A discussion of the strength of the Friedel oscillations in a realistic quasi 
one-dimensional Fermi gas with interactions is addressed at the end of the 
Appendix.  
 
\section{One--Particle Momentum Distributions}

Even for a confined system it makes sense to study the momentum density 

\begin{equation}\label{3.1}
p(k) \equiv \langle \hat{c}^+ _{k} \hat{c}_k \rangle.
\end{equation}

The operator $\hat{c}_k$ annihilates a fermion with (continuous) momentum 
$\hbar k$. It can be decomposed into the fermionic annihilation operators 
of the harmonic oscillator according to (\ref{A.6}).

Taking into account that the expectation values $\langle \hat{c}^+ _m  \hat{c}_n \rangle $
vanish unless $|m-n|$ is even, the momentum density becomes

\begin{eqnarray}\label{3.4}
p(k) &=& \sum _{m,n = 0}^\infty ( f_m^k )^* f_n^k \,(-1)^{m+n}\,
\langle \hat{c}^+ _m \hat{c} _n \rangle \\[4mm]\nonumber
&=&\frac{1}{\alpha^2}\,\sum _{m = 0}^\infty\sum _{p=-m}^m (-1)^p \psi _{m-p} 
(\frac{k}{\alpha^2})
\psi _{m+p} (\frac{k}{\alpha^2})\,\langle \hat{c}^+ _{m-p} \hat{c} _{m+p} \rangle.
\end{eqnarray}

Obviously, the momentum distribution always satisfies the sum rule

\begin{equation}\label{3.5}
\int _{-\infty}^{\infty} dk\,p(k) = N.
\end{equation}

If the expectation values (\ref{2.11}) were diagonal, as in 
the non--interacting case, one would obtain the relation

\begin{eqnarray}\label{3.6}
p(k) \equiv \frac{1}{\alpha^2}\,n( z = k/\alpha^2). 
\end{eqnarray}

However, the non-diagonal matrix elements are relevant: 
From (\ref{2.8}) it is seen that in case of IM1 the matrix elements obey the
relation $\langle \hat{c}^+ _{M-p}\hat{c} _{M+p} \rangle(-\alpha _1)
=(-1)^p \langle \hat{c}^+ _{M-p}\hat{c} _{M+p} \rangle(\alpha _1)$ leading 
to the remarkable result 

\begin{eqnarray}\label{3.7}
p(k, \alpha _1) = \frac{1}{\alpha^2}\,n( z = k/\alpha^2,-\alpha _1),
\end{eqnarray}

for IM1, i.e., the momentum distribution functions behaves oppositely to the 
density: Attractive interactions increase the oscillations (with period $\pi 
\alpha^2/k_F$), repulsive interactions decrease them. This effect is also found 
for IM2 though it is less pronounced than in the density \cite{REM1}. 

\section{Summary}

Using methods borrowed from the theory of the Luttinger model, we propose
a scheme to treat interacting fermions in a one--dimensional harmonic
trap analytically. As in the Luttinger model, the method rests on the
linear dispersion of non--interacting fermion states which is extended to
negative energies. A crucial step in the treatment is the correspondence between 
fermionic operators and exponentials of bosonic fields. This allows to study
the fermion density distribution in the trap. A detailed investigation of the
relation between the model interaction and realistic pair interactions is also
given.

{\bf Acknowledgements}: The author thanks G. Alber, S. N. Artemenko, F. Gleisberg, 
F. Lochmann, T. Pfau, U. Schl\"oder, W. Schleich, J. Voit, and C. Zimmermann for 
valuable discussions and Deutsche Forschungsgemeinschaft for financial support.

\section{Appendix}

\newcounter{affix}
\setcounter{equation}{0}
\setcounter{affix}{1}
\renewcommand{\theequation}{\Alph{affix}.\arabic{equation}}

The Appendix compiles a number of formulae relating a physical two--body interaction
to the four fermion matrix element in (\ref{1.16}). We assume a local and translation 
invariant pair interaction

\begin{eqnarray}\label{A.0}
V(z_1,z_2;z_3,z_4)=V(z_1-z_2)\,\delta(z_1-z_3)\,\delta(z_2-z_4),
\end{eqnarray}

and start from the representation of the interaction operator in the momentum basis:

\begin{eqnarray}\label{A.1}
\hat{V} = \frac{1}{4\pi} \int^\infty _{-\infty} \, dk_1 \, dk\,dk'\,  \tilde{V}(k_1) 
\hat{c}^+ _{k+k_1}\hat{c}^+ _{k'-k_1} \hat{c}_{k'} \hat{c}_k.
\end{eqnarray}

The Fourier transform of the interaction potential is 

\begin{eqnarray}\label{A.2}
\tilde{V} (k) = \int^\infty _{-\infty} \, dz \, e^{ikz} \,V(z) 
= \tilde{V} (-k)= \tilde{V}^*(k),
\end{eqnarray}

and the  exchange symmetry of the pair potential $V(z) = V(-z)$ has been utilized.

By rearranging the operators in (\ref{A.1}), one finds

\begin{eqnarray}\label{A.4}
\hat{V} = \frac{1}{4 \pi} \int^\infty _{-\infty} \, dk_1 \, dk\,dk' \, \tilde{V}(k_1) 
\left( \hat{c}^+ _{k+k_1} \hat{c}_k \right)\left( \hat{c}^+ _{k'-k_1} \hat{c}_{k'} \right)
 - \frac{1}{2} \hat{N} V (z=0),
\end{eqnarray}   

where $\hat{N}$ is the fermion number operator.
The self--energy contribution in (\ref{A.4}), which is finite for any reasonably 
regularized potential, will be omitted. We go from the momentum basis to the 
harmonic oscillator basis by means of

\begin{eqnarray}\label{A.6}
\hat{c}_k = \sum ^\infty _{n=0} (-1)^n \, f^k_n \, \hat{c}_n,
\end{eqnarray}

with \cite{GWSZ00}

\begin{eqnarray}\label{A.7}
f^k_n \equiv \frac{1}{\sqrt{2\pi}} \, \int ^ \infty _ {-\infty} \, dz \, e^{ikz} \psi _n (z)
=\frac{i^n}{\alpha} \psi _n (z =\frac{k}{\alpha ^2}). 
\end{eqnarray}

Thus the usual form for the matrix elements, as used in the interaction operator

\begin{eqnarray}\label{A.8}
\hat{V} = \frac{1}{2}\,\sum ^\infty _{mnpq=0}  V (m,p;q,n)\, ( \hat{c}^+_m \hat{c}_q) 
(\hat{c} ^+ _p \hat{c}_n),
\end{eqnarray}

namely

\begin{eqnarray}\label{A.9}
V(m,p;q,n) =\int ^\infty _{-\infty} \, dz_1 dz_2\, \psi _m (z_1) \psi _p (z_2)
V (z_1- z_2) \psi  _q (z_1) \psi _n (z_2),
\end{eqnarray}

(with the obvious symmetries $m \leftrightarrow q$, $n \leftrightarrow p$) 
takes on the less familiar form

\begin{eqnarray}\label{A.10}
V(m,p;q,n) =(-1)^{n+q} \frac{1}{2 \pi} \int^\infty _{-\infty} \, dk_1 \,dk \,dk' \,
\tilde{V}(k_1) \left\{ f_m^{k-k_1} f_q^k \right\} \left \{ f_p^{k'+k_1} f^{k'}_n \right \}.
\end{eqnarray}

It can be further evaluated by performing the convolution in (\ref{A.10}). By using 
the method described in \cite{GWSZ00}, one obtains

\begin{eqnarray}\label{A.11}
V(m,p;q,n) = (-1)^{n+q}\, i ^{m+n+p+q}
\frac{1}{2\pi} \, \int ^\infty _{-\infty} \, dk_1\, C_{mq} (k_1) \tilde{V}(k_1) C_{pn}(-k_1),
\end{eqnarray}

with ($q \ge m$ is assumed)

\begin{eqnarray}\label{A.12}
C_{mq} (k_1) &\equiv& \frac{1}{\alpha ^2} \int^\infty _{-\infty} \, dk\, 
\psi _m \left( \frac{k-k_1}{\alpha ^2} \right)
 \psi _q \left( \frac{k}{\alpha ^2} \right)
 \\[4mm]\nonumber
&=& \sqrt{ \frac{m!}{q!}} \, \left( \frac{k_1}{\alpha\sqrt{2}} \right ) ^{q-m}
\exp \left[-\frac{k^2_1}{4 \alpha ^2}\right]\, L^{(q-m)}_m  \left( \frac{k^2_1}
{2\alpha ^2} \right )=(-1)^{m+q}\,C_{qm}(k_1).
\end{eqnarray}

In order to eliminate a possible contact interaction in $V(z)$, we set

\begin{eqnarray}\label{A.13}
V_{\mbox{eff}}(k)\equiv \tilde{V}(k)-\tilde{V}(k=\infty),
\end{eqnarray} 

to obtain the following exact expression for the matrix elements of the potential
in the basis of oscillator states and for $q \ge m$, $n \ge p$:

\begin{eqnarray}\label{A.14}
&&V(m,p;q,n)=i^{m+n+p+q}\{(-1)^{q-p}+(-1)^{n-m}\}\sqrt{\frac{m! p!}{2\,n! q!}}
\,\frac{\alpha}{2 \pi}\\[4mm]\nonumber
&&\int _0^\infty dv\,e^{-v}v^{(q-m+n-p-1)/2}\,V_{\mbox{eff}}(k=\alpha \sqrt{2 v})\,
L_m^{(q-m)}(v)\, L_p^{(n-p)}(v). 
\end{eqnarray} 

These matrix elements are of particular importance for values of the
indices $m$, $n$, etc., which correspond to wave functions near the Fermi level, i.e., 
$m \approx N$ etc.. Setting $q=m+Q$ and exploiting $m \gg Q$ (e.g., $Q \le \sqrt{N})$,
the asymptotic expansion \cite{AS70} of the Laguerre polynomials can be applied:

\begin{eqnarray}\label{A.15}
L_m^{(Q)} (v) \sim \left( \frac{m}{v} \right )^{\frac{Q}{2}} 
J_Q(2 \sqrt{m v}).
\end{eqnarray}

This formula implies the approximation $(m+Q)! \approx m^Q m!$. Setting 
 $n=p+R$, one finds

\begin{eqnarray}\label{A.15 a}
V(m,p;m+Q,p+R) \sim &&i^{Q+R}\frac{(-1)^Q+(-1)^R}
{\sqrt{2}}\, \frac{\alpha}{2 \pi}\,
\\[4mm]\nonumber
&&  \int^ \infty _0 \frac{d v}{\sqrt{v}}\, e^{-v}\,V_{\mbox{eff}}\left(k=\alpha \sqrt{2 v}
\right)\,J_Q(2 \sqrt{m v})\,J_R(2 \sqrt{p v}).
\end{eqnarray}

The integration region near $v=0$ will contribute little to matrix elements with 
$Q >0$ and $R>0$ (the $Q=R=0$ case in (\ref{1.17},\ref{1.19}) is irrelevant)
provided $\tilde{V}(k)$ is bounded
for small wave numbers $k$. In order to estimate these matrix elements, we can thus assume
$v$ to be not less than of order unity and apply the asymptotic expansion of the Bessel
functions $J$. This gives: 

\begin{eqnarray}\label{A.16}
V(m,p;m+Q,p+R) &&\approx i^{Q+R} \frac{(-1)^Q+(-1)^R}
{\sqrt{2}}\, \frac{\alpha}{\pi^2}\, \int^ \infty _a
\frac{du}{u} e^{-\frac{u^2}{4}}
\\[4mm]\nonumber
&& \cos \left( u \sqrt{m}- \frac{Q }{2} \pi - \frac{\pi}{4} \right)
V_{\mbox{eff}}\left(k=\frac{\alpha u}{\sqrt{2}}\right)
\cos \left( u \sqrt{p}- \frac{R \pi}{2} - \frac{\pi}{4} \right),
\end{eqnarray}               

where $a$ is of order unity. We choose $a=1$. Finally, setting $R=Q+\Delta R$, 
leads to 

\begin{eqnarray}\label{A.17}
V(m,p;m+Q,p+Q+\Delta R) &&\approx  \frac{\alpha}
{\pi^2 \sqrt{2}}\, \cos(\frac{\pi\Delta R}{2})\, \int^ \infty _1
\frac{du}{u} e^{-\frac{u^2}{4}}V{\mbox{eff}}(k=\frac{\alpha u}{\sqrt{2}})
\\[4mm]\nonumber
&& \left \{\cos \left( u (\sqrt{m}- \sqrt{p})\right)+(-1)^Q
\sin \left( u (\sqrt{m}+ \sqrt{p}) \right)\right\}.
\end{eqnarray}

Equation (\ref{A.17}) admits the following discussion of the matrix elements:

\begin{itemize}
\item[1.] The matrix elements depend weakly on $m$ and $p$ in the ranges $|\Delta m|
\equiv |m-N|< \sqrt{N}$ and $|\Delta p| \equiv |p-N| <\sqrt{N}$.
\item[2.] Elements with $\Delta R=0$ are (nearly) independent of $Q < \sqrt{N}$ and
all have the same sign: They are positive for repulsive interactions and negative 
in the opposite case. 
\item[3.] Varying $\Delta R$, produces alternating signs which weakens the effect 
of these matrix elements. 
\end{itemize}

These are essentially the properties which justify the model interactions
(\ref{1.17}) and (\ref{1.19}). In this connection, it should be mentioned that, as
in Luttinger liquid theory, one does not need a faithful representation of the physical
interaction \cite{H81,VM99} to capture essential features of the interaction.

For a numerical evaluation of the interaction strength,
we begin with the interaction of longitudinally aligned magnetic
dipoles (cf. \cite{GRP00,GER01}) in a quasi one--dimensional magnetic trap for which the 
effective (attractive) potential is

\begin{eqnarray}\label{A.18}
V(z) = - \frac{\mu _0}{2 \pi} \mu ^2 \frac{1}{(z^2+d^2)^{3/2}}.
\end{eqnarray}

Here, an ultra--violet cut off has been introduced. It is given by the
width of the one--dimensional channel, i.e., $d=\alpha _t^{-1}$, where $\alpha _t$
is the reciprocal harmonic oscillator length of the transverse (ground state) wave
function. This length is large on an atomic scale. Introducing the ratio
$\lambda  \equiv \omega _l/\omega _t$, one has $d=\sqrt{\lambda}/\alpha$. In case of a 
completely filled one--dimensional trap, which we henceforth assume, $N=1/\lambda$ holds. 

Fourier transforming the potential (\ref{A.18}), gives

\begin{eqnarray}\label{A.19}
\tilde{V} (k) = V_0 ( \sqrt{\lambda}\, \frac{k}{\alpha}) 
K_1 (\sqrt{\lambda}\, \frac{k}{\alpha}). 
\end{eqnarray}

This expression involves a modified Bessel function
and is bounded (though not analytic) at $k=0$. It vanishes for large momenta.
The prefactor contains the magnetic moment $\mu$ of the fermions and reads: 

\begin{eqnarray}\label{A.20}
V_0 = - \frac{\mu _0}{\pi}\,  \frac{\mu^2 \alpha ^2}{\lambda}.
\end{eqnarray}

For IM1, the quantity

\begin{eqnarray}\label{A.21}
V(1) \approx \left[V_0 \frac{\alpha}{\pi^2 \sqrt{2}}\right]\, \int^\infty _1 \, du\,
\{\sqrt{\frac{\lambda}{2}}\,K_1 (u \sqrt{\frac{\lambda}{2}})\}\,\exp\{- \frac{u^2}{4}\}
 \, \cos \left(u (\sqrt{m}- \sqrt{p})\right)
\end{eqnarray}

is needed. The contribution from the second term in (\ref{A.17}) containing 
$\sqrt{m}+ \sqrt{p}$ is negligible. The factor in square brackets can be written 
as 

\begin{eqnarray}\label{A.22}
\left[V_0 \frac{\alpha}{\pi^2 \sqrt{2}}\right]= -\frac{1}{\sqrt{2}}\left(\frac{\mu _0 
\mu^2 m_A \alpha N}{\pi^3 \hbar^2}\right)\,\hbar \omega _\ell,
\end{eqnarray}

where $m_A$ is the atom mass.
The prefactor of the scaling energy $\hbar \omega _\ell$ is estimated for $^6 Li$ 
to be $-3 \cdot 10^{-3}(N/10^4)$ adopting the data from \cite{GWSZ00}. The remaining integral
in (\ref{A.21}) is about unity for $|m-p| < \sqrt{N}$. So it needs about
$N=10^6$ atoms in the trap to produce a significant effect of the dipole--dipole interaction.

For scattering processes in three dimensions, the van der Waals potential is more relevant. 
The dilute atoms
in the trap (mean distance larger than 100 Bohr radii) explore mostly the long--range
part of it. Reducing the $R^{-6}$ potential to an effective potential in one dimension,
gives: 

\begin{eqnarray}\label{A.23}
V(z) = - \frac{A}{(z^2+d^2)^3},
\end{eqnarray}

with the Fourier transform

\begin{eqnarray}\label{A.24}
\tilde{V} (k) = - \frac{\pi}{8} \frac{A}{d^5}\, e^{-kd}\, (3+3kd+k^2d^2).
\end{eqnarray}

Following \cite{ZS93,CDJ94}, the coefficient $A$ can be expressed in terms of the 
coefficient $C_6$ therein, giving

\begin{eqnarray}\label{A.25}
V(1) \approx \left[\left(\frac{C_6}{16 \pi \sqrt{2}}\right)\,e_0^2\,a_0^5\,
\alpha _t^5\,\alpha \right]\,&&\hbar \omega _\ell\, \int^\infty _1 \,\frac{du}{u}\,
\exp(-u^2/4 -u \sqrt{\lambda/2})
\\[4mm]\nonumber
&&\left(3+3u \sqrt{\lambda/2} +u^2 \lambda/2\right)\,
 \, \cos \left(u (\sqrt{m}- \sqrt{p})\right),
\end{eqnarray}

where $e_0$ and $a_0$ are electron charge and Bohr radius, respectively.
Fixing $\omega _\ell$ at $2 \pi \cdot 10 s^{-1}$, as in \cite{GWSZ00}, 
the evaluation of (\ref{A.25}) gives
a total prefactor of $\hbar \omega _\ell$ according to $\approx -6 \cdot 
10^{-7}\,(N/10^4)^{5/2}$. Even for $N=10^6$, this is small compared to the
dipole--dipole interaction. This reversal in the importance of the two 
interactions can be traced back to the ultra--violet cutoff $d \equiv 
\alpha _t^{-1}$ needed in the quasi one--dimensional forms of the interactions. 

The choice of transverse cut-off determines the strength of the effective 
one-dimensional interaction. Our choice is motivated by the following fact: The 
transverse trap direction is associated with only one intrinsic length scale $(L_F)_t
=1/\alpha _t = l_t =l \sqrt{\lambda} \approx 1/k_F$ and this length coincides with the
shortest intrinsic length scale of the longitudinal trap direction. 

The amplitude of the Friedel oscillations inside a bounded Fermi sea scales with 
the particle number $N$ as $1/N$. Adopting the numbers estimated above, it is impossible 
to detect Friedel oscillations in a gas of $10^6$ $^6 Li$ atoms with present techniques. 
However, the interaction strength $V(1)$ between magnetic 
dipoles scales as $\mu^2 m_A^{3/2}$. Using $^{53} Cr$ instead of $^6 Li$ enhances the 
interaction by nearly three orders of magnitude. Fermionic molecules are even more 
promising, especially when they are polar (cf., e.g., \cite{BBC00}): An electric dipole 
moment of 1 Debye ($p=a_0 e_0/2.5$) leads to an interaction strength which is enhanced 
by a factor $(0.8/\alpha _S)^2 \approx 10^5$ ($\alpha _S$: fine structure constant) over 
that from magnetic dipoles of magnitude $\mu _B$ ($\mu _B$: Bohr's magneton).

Thus it seems that the issue of Friedel oscillations in an interacting quasi 
one-dimensional degenerate Fermi gas is not purely academic.


\begin{references}
\bibitem{REM1} An approximate analytical evaluation of (\ref{2.11}) is
possible for negative $\alpha _0$. By setting $s=0$ in the last factor
of the integrand in (\ref{2.11}), the integration over $t$ becomes trivial,
giving $\delta _{p,0}$, i.e., the matrix elements become diagonal. The integration 
over $s$ leads to hypergeometric functions in two variables which are difficult 
to discuss.This approximation to the matrix elements is reasonable for calculating 
the density in the attractive case. However, the contributions of the (small) 
non-diagonal matrix elements to the momentum distribution interfere constructively 
enhancing the oscillations.
\end{references}
\end{document}